\documentclass[aps,pra,amsmath,amsfonts,amssymb,twocolumn,nofootinbib]{revtex4}

\usepackage{amssymb}
\usepackage{graphicx,psfrag,color}
\usepackage{dcolumn}
\usepackage{bm}
\usepackage{mathbbol}
\usepackage[T1]{fontenc}

\newcommand{\med}[1]{\left\langle #1 \right\rangle}

\begin{document}
\flushbottom

\title{Detecting quantum critical points at finite temperature via quantum teleportation: further models}

\author{G. A. P. Ribeiro}
\author{Gustavo Rigolin}
\email{rigolin@ufscar.br}
\affiliation{Departamento de F\'isica, Universidade Federal de S\~ao Carlos, 13565-905, S\~ao Carlos, SP, Brazil}

\date{\today}

\begin{abstract}
In [Phys. Rev. A \textbf{107}, 052420 (2023)] we showed that the quantum teleportation protocol can
be used to detect quantum critical points (QCPs) associated with a couple of different classes of quantum phase transitions, even when the system is away from the absolute zero temperature ($T=0$). 
Here, working in the thermodynamic limit (infinite chains), we extend the previous analysis for several other spin-1/2 models. We investigate the usefulness of the quantum teleportation protocol to 
detect the QCPs of those models when the temperature is either zero or greater than zero. The spin chains we investigate here are described by the XXZ model, the XY model, and the Ising model, all of them subjected to an external magnetic field. 
Specifically, we use a pair of nearest neighbor qubits 
from an infinite spin chain at thermal equilibrium with a reservoir at temperature 
$T$ as the resource to execute the quantum teleportation protocol. We show that the 
ability of this pair of qubits to faithfully teleport an external qubit from the chain is dramatically affected as we cross the QCPs related to the aforementioned models. 
The results here presented together with the ones of [Phys. Rev. A \textbf{107}, 052420 (2023)] suggest that the quantum teleportation protocol is a robust and quite universal tool to detect QCPs
even when the system of interest is far from the absolute zero temperature.
\end{abstract}


\maketitle

\section{Introduction} 

A quantum phase transition (QPT) is a qualitative change in the ground state
of a many-body system that theoretically 
happens at the absolute zero ($T=0$) as we slowly change the 
system's Hamiltonian $H$ \cite{sac99,gre02,gan10,row10}. 
This qualitative change in the physical properties of the system is driven 
by genuine quantum fluctuations (Heisenberg 
uncertainty principle) since at $T=0$ there are no thermal fluctuations
at stake. A QPT is usually characterized by a symmetry change in the 
system's ground state and by the emergence of an order parameter such as
the total magnetization that is no longer zero after a ferromagnetic QPT.

Most of the theoretical analysis studying QPTs, in particular those employing
quantum information theory concepts, assume that the system is 
at $T=0$ \cite{wu04,oli06,dil08,sar08}.
Experimentally, though, 
we cannot cool a many-body system to $T=0$ (third law of thermodynamics) and thus
it is crucial to build and develop robust tools to characterize QPTs assuming the
system is at finite $T$. This becomes even more important whenever 
$kT\approx \Delta E$, where $k$ is Boltzmann constant and 
$\Delta E$ is the energy gap between the system's ground and first excited states. In this scenario thermal
fluctuations cannot be ignored and it is a necessity to develop
robust quantum critical point (QCP) detectors that still work in this regime.
For instance, the entanglement 
of formation (EoF) \cite{woo98}
and the magnetic susceptibility no longer detect a QPT in spin chains
when $T>0$ and other tools are needed to detect a QCP at finite $T$ \cite{wer10,wer10b}.

A very useful and robust tool to detect QCPs at finite $T$ is quantum discord (QD),
usually called thermal quantum discord (TQD) in this context \cite{wer10b}. Although very successful in
detecting QCPs when $T>0$ \cite{wer10b}, QD \cite{oll01,hen01} has its handicaps. 
The computation of QD is NP-complete \cite{hua14}, which implies that the 
evaluation of QD is an intractable problem for systems described by a large Hilbert space \cite{mal16}. Also, QD has no operational interpretation. 
We do not have an experimental procedure to directly measure QD. 
We can only compute QD if we have access to the system's whole density matrix.

We should note that, recently, a quantity derived from the quantum coherence \cite{wig63,fan14,gir14} was shown to be very robust to detect QCPs using finite $T$ data, outperforming QD for certain 
models \cite{li20}. This quantity was called the logarithm of the spectrum 
of the quantum coherence ($L_{QC}$) \cite{li20}.  
But similarly to QD, $L_{QC}$ has no operational interpretation, i.e., there is no experimental procedure for its direct determination. One needs the density matrix (measured or calculated) of the system investigated to compute it. For a two-qubit density matrix $\rho$, this means that we always need to know (compute or measure) its one- and two-point correlation functions. 
Furthermore, to compute $L_{QC}$ one needs the 
eigenvalues of the following squared commutator, 
$[\rho, K]^2$, whose computational complexity 
does not scale linearly with the size of the 
system as we increase its Hilbert space dimension. 
The computation of $L_{QC}$ has also an arbitrariness in the choice of the 
observable ``K'' \cite{li20}. 
Depending on the observable chosen, $L_{QC}$ does not detect QCPs. And for high dimensional systems, the number of observables becomes very large, making it difficult to test all cases and increasing the arbitrariness for the choice
of the right observable. 

In Ref. \cite{pav23} we developed a QCP detection tool that has the most useful
characteristics of TQD in spotlighting QCPs at finite $T$ and, 
in addition, is free from the handicaps outlined above.    
That tool is built on the quantum teleportation 
protocol \cite{ben93,yeo02,rig17,rig15} and will be described in Sec. \ref{tool}. We should also mention another recent tool to 
detect QCPs at $T=0$ based on the quantum energy teleportation protocol
\cite{hot08,hot09}. It was show in Refs. \cite{ike23a,ike23b,ike23c,ike23d}
that for several models the amount of teleported energy  depends on 
the phase of the system. 

In this work we apply the teleportation based QCP detector of Ref. \cite{pav23} to
several other models. Here we study the XXZ model subjected to an external magnetic field, complementing the analysis of Ref. \cite{pav23}, where we studied this model 
without an external field. We also investigate the efficiency of the teleportation 
based QCP detector in spotlighting the QCPs of 
the Ising model and of the XY model in a transverse magnetic field. As we will
see, the present tool allows us to determine all the QCPs of these models even if
the system's temperature is not zero.

\section{The teleportation based critical point detector}
\label{tool}

\begin{figure}[!ht]
\includegraphics[width=7.5cm]{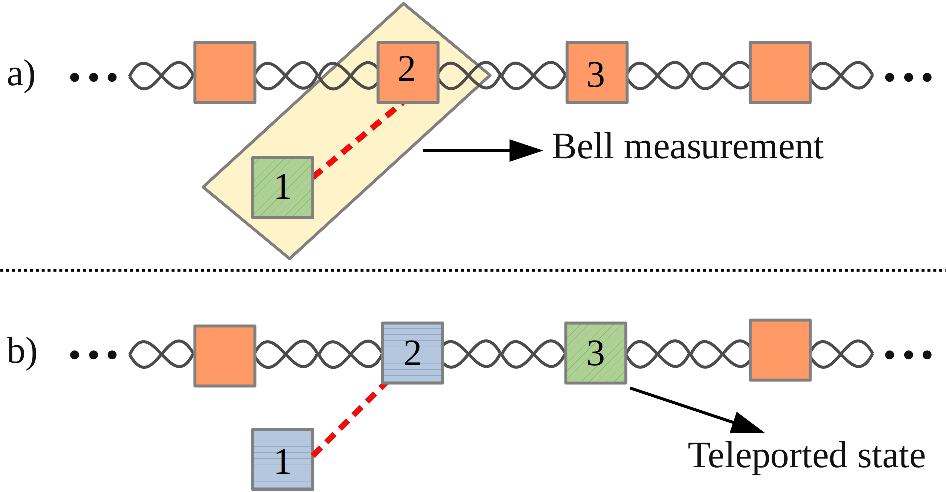}
\caption{\label{fig_scheme}(color online) 
a) Alice and Bob agree that spins 2 and 3 of the spin chain are the entangled resource
used to teleport the external qubit 1. Alice implements a Bell measurement (BM)  
onto qubits 1 and 2.
b) Alice tells Bob her BM result via a
classical communication channel. Bob then applies a 
unitary operation on qubit 3, depending on the news received from Alice. 
This finishes one run of the protocol.}
\end{figure}

Let us start by reviewing the standard teleportation protocol \cite{ben93}, in 
particular its mathematical description when the shared entangled state between
Alice and Bob is not a pure state \cite{rig17,rig15}. We label the qubits
from the entangled resource shared by Alice and Bob as qubits $2$ and $3$,
respectively (see Fig.~\ref{fig_scheme}). 
The density matrix describing those qubits is $\rho_{23}$. 
The qubit that Alice wants to teleport to Bob
is a pure state external to the spin chain and its density matrix is 
$\rho_1=|\psi\rangle\langle \psi|$ (qubit $1$ in Fig.~\ref{fig_scheme}),
where
\begin{equation}
|\psi\rangle
=r|0\rangle+\sqrt{1-r^2}e^{i\chi}|1\rangle
=\cos(\theta/2)|0\rangle
+\sin(\theta/2)e^{i\chi}|1\rangle,
\label{step0}
\end{equation}
with $0\leq r \leq 1 (0\leq \theta \leq \pi)$ and $0\leq \chi < 2\pi$. 

At the beginning of the 
teleportation protocol, the state describing the three qubits is 
\begin{equation}
 \rho = \rho_{1} \otimes \rho_{23}.
\label{stepA}
\end{equation}
At the end of the protocol (after one run of the protocol), Bob's spin (qubit 3) is given by \cite{pav23,rig17}
\begin{equation}
\rho_{_{B_j}}=   \frac{U_jTr_{12}[P_j \rho P_j]U_j^\dagger}{Q_j(|\psi\rangle)}.
\label{stepD}
\end{equation}
Here $Tr_{12}$ is the partial trace on Alice's spins (qubits 1 and 2), 
$j$ denotes the Bell measurement (BM) result
obtained by Alice ($j=\Psi^-,\Psi^+,\Phi^-,\Phi^+$), and $P_j$ 
represents the four projectors describing the BMs,
\begin{eqnarray}
P_{\Psi^{\pm}} &=& |\Psi^{\pm}\rangle \langle \Psi^{\pm}|, \label{projectorA}\\  
P_{\Phi^{\pm}} &=& |\Phi^{\pm}\rangle \langle \Psi^{\pm}|, \label{projectorB}  
\end{eqnarray}
with the Bell states given by
\begin{eqnarray}
|\Psi^{\mp}\rangle&=&(|01\rangle \mp |10\rangle)/\sqrt{2},\label{BellA} \\ 
|\Phi^{\mp}\rangle&=&(|00\rangle \mp |11\rangle)/\sqrt{2}. \label{BellB}
\end{eqnarray}

The probability to measure a given Bell state $j$ is \cite{pav23,rig17}
\begin{equation}
Q_j(|\psi\rangle) = Tr[{P_j \rho}]
\label{prob}
\end{equation}
and the unitary correction that Bob should implement on his qubit after receiving the news about Alice's BM result is $U_j$. 

The unitary operation that Bob should apply on his qubit at the end of a given 
run of the protocol also depends on the entangled state shared with Alice. 
When they share a maximally entangled pure
state $|k\rangle$ (Bell state) \cite{ben93}, the set $S_k$ below lists the 
four unitary operations that Bob should apply on his qubit \cite{pav23,rig17},
\begin{eqnarray}
S_{\Phi^+}=\{U_{\Phi^+},U_{\Phi^-},U_{\Psi^+},U_{\Psi^-}\}
=\{\mathbb{1},\sigma^z,\sigma^x,\sigma^z\sigma^x\},
\label{s1} \\
S_{\Phi^-}=\{U_{\Phi^+},U_{\Phi^-},U_{\Psi^+},U_{\Psi^-}\}
=\{\sigma^z,\mathbb{1},\sigma^z\sigma^x,\sigma^x\}, \label{s2}\\
S_{\Psi^+}=\{U_{\Phi^+},U_{\Phi^-},U_{\Psi^+},U_{\Psi^-}\}=\{\sigma^x,\sigma^z\sigma^x,\mathbb{1},\sigma^z\}, \label{s3}\\  
S_{\Psi^-}=\{U_{\Phi^+},U_{\Phi^-},U_{\Psi^+},U_{\Psi^-}\}=\{\sigma^z\sigma^x,\sigma^x,\sigma^z,\mathbb{1}\},\label{s4}
\end{eqnarray}
where $\mathbb{1}$ is the identity matrix and $\sigma^\alpha$, $\alpha=x,y,z$,
is the standard Pauli matrix \cite{nie00}. In other words,
$S_k$ represents the set of unitary operations that Bob should apply if Alice and Bob share the Bell state $|k\rangle$,
with $k=\Psi^\pm,\Phi^\pm$. For instance, $S_{\Phi^+}$ means that they share the 
state $|\Phi^+\rangle$ and that if Alice's BM result is $|\Phi^+\rangle,|\Phi^-\rangle,|\Psi^+\rangle$, or $|\Psi^-\rangle$, the corresponding unitary corrections
that Bob should apply is $\mathbb{1},\sigma^z,\sigma^x$, or $\sigma^z\sigma^x$.

In the models we will be studying in what follows, the state $\rho_{23}$ shared 
between Alice and Bob is a mixed state. In one quantum phase $\rho_{23}$
is closer to one of the four Bell states and in another phase closer to a different 
one. Thus, when studying the QCPs of a spin chain we will employ the four sets
of unitary operations above, eventually picking the set yielding the optimal teleportation protocol.  

To determine the optimal teleportation protocol, we need a quantitative measure of the similarity between the teleported state at the end of a run of the protocol and the initial state teleported by Alice. As usual, we employ the fidelity \cite{uhl76} to quantify the similarity between those states. When we have a pure input state
the fidelity is
\begin{equation}
F_j(|\psi \rangle,S_k) = \langle \psi | \rho_{_{B_j}} | \psi \rangle,
\label{Fidj}
\end{equation}
where $| \psi \rangle$ is given by Eq.~(\ref{step0}) and $\rho_{_{B_j}}$ by
Eq.~(\ref{stepD}). 
Note that the subscript $j$ denotes which Bell state Alice obtained after
implementing the BM on qubits 1 and 2. For a teleported state exactly equal to 
the input state we have $F_j=1$, while $F_j=0$ if the teleported state 
is orthogonal to the input. We should note that in addition to depending on
the initial state, $F_j$ also depends through $\rho_{_{B_j}}$ on the entangled state shared by Alice and Bob and on the set of unitary operations $S_k$ that he can apply
on his qubit. In this work, the entangled resource is determined by the model being investigated and we can freely choose $|\psi \rangle$ and $S_k$, with $k=\Psi^\mp,\Phi^\mp$.

If we fix the input state, after a single run of the protocol its fidelity is given by Eq.~(\ref{Fidj}) and after several runs of the protocol the mean fidelity (efficiency) is \cite{pav23,rig17,rig15,gor06}
\begin{equation}
\overline{F}(|\psi\rangle,S_k)= \sum_{j=\Psi^{\mp},\Phi^{\mp}} 
Q_j(|\psi\rangle)F_j(| \psi \rangle,S_k). \label{Fbar}
\end{equation}
Equation (\ref{Fbar}), as we show here, is the building block leading to the 
most sharp QCP detector and can be understood as the efficiency of the teleportation protocol for a fixed input state and a given set $S_k$ of unitary operations.

In order to obtain an input state independent measure of the efficiency of the quantum teleportation protocol, we take the average over all states on the Bloch sphere. 
This Bloch sphere average is equivalent to assuming in Eq.~(\ref{step0}) that
$r^2$ and $\gamma$ are two independent continuous random variables over their domain \cite{rig15,gor06}. We can write this state independent mean fidelity as
\cite{gor06,rig15,rig17}
\begin{equation}
\langle \overline{F}(S_k) \rangle = \int_\Omega \overline{F}(|\psi\rangle,S_k) 
\mathcal{P}(|\psi\rangle)d|\psi\rangle.
\label{Flangle}
\end{equation}
In Eq.~(\ref{Flangle}) the integration over the sample space $\Omega$ includes 
all qubits on the Bloch sphere and $\mathcal{P}(|\psi\rangle)$ is the appropriate
uniform probability distribution over $\Omega$ \cite{pav23}. 
From now on, the quantity defined in Eq.~(\ref{Fbar}) will be called  
``mean fidelity'' and the quantity given by Eq.~(\ref{Flangle}) will be denoted
``average fidelity''.

\section{The XXZ model in an external field}
\label{secXXZ}

The Hamiltonian describing the XXZ model in an external longitudinal field is ($\hbar=1$)
\begin{equation}
H = \sum_{j=1}^{L}\left(\sigma^{x}_{j}\sigma^{x}_{j+1} +
\sigma^{y}_{j}\sigma^{y}_{j+1} + \Delta
\sigma^{z}_{j}\sigma^{z}_{j+1} - \frac{h}{2}\sigma^z_j\right). \label{Hxxz}
\end{equation}
We will be dealing with a spin-1/2 chain in the thermodynamic
limit ($L\rightarrow \infty$) satisfying periodic boundary conditions
($\sigma^{\alpha}_{L+1} = \sigma^{\alpha}_1$). The subscript $j$ above means that
$\sigma^{\alpha}_j$ acts on the spin at the lattice site $j$.
The anisotropy $\Delta$ is our tuning parameter and $h$ is the external magnetic field, which will be fixed as we vary $\Delta$ across the QCPs for this model.

At $T=0$  and for a 
finite external magnetic 
field $h$, this model has two
QCPs \cite{yan66,clo66,klu92,bor05,boo08,tri10,tak99}. At $\Delta_1$ we have the first QCP,
where the ground state changes from a ferromagnetic ($\Delta < \Delta_1$) to a critical 
antiferromagnetic phase ($\Delta_1 < \Delta < \Delta_{2}$). 
At $\Delta_{2}$ 
another phase
transition takes place, 
with the system becoming an Ising-like antiferromagnet for 
$\Delta > \Delta_{2}$. The two QCPs depend on $h$ and are given as
follows \cite{yan66,clo66,klu92,bor05,boo08,tri10,tak99}. 

The critical point $\Delta_1$
is obtained by solving the 
following equation once we fix the value of $h$,
\begin{equation}
h=4J(1+\Delta_1).
\label{d1}
\end{equation}
The critical point 
$\Delta_2$
is the solution of 
\begin{equation}
h=4\sinh(\eta)\sum_{j=-\infty}^\infty\frac{(-1)^j}{\cosh(j\eta)},
\label{dinf}
\end{equation}
where $\eta = \cosh^{-1}(\Delta_{2})$.
In Table \ref{d1dinf} we list the QCPs for the two values of $h$ that we will
be dealing here and also the QCPs for the zero field case ($h=0$).
\begin{table}[!ht]
\caption{\label{d1dinf}
Quantum critical points for different values of the external 
field $h$. The values for $\Delta_{2}$ when $h>0$ 
are accurate within an error of $\pm 0.001$.} 
\begin{ruledtabular}
\begin{tabular}{lrrr}
& $h=0$ & $h=6$ & $h=12$\\ \hline 
$\Delta_1$ & -1.00 & 0.50 & 2.00\\ 
$\Delta_{2}$  & 1.00 & 3.299 & 4.875\\
\end{tabular}
\end{ruledtabular}
\end{table} 
 
A physical system in equilibrium with a thermal reservoir at temperature $T$ is 
described by the canonical ensemble density matrix. As such, 
the density matrix describing the thermalized spin chain  
(\ref{Hxxz}) is $\varrho=e^{-H/kT}/Z$, where 
$Z=Tr[e^{-H/kT}]$ is the partition function and $k$ is Boltzmann's constant.
To obtain the density matrix describing a pair of nearest neighbor spins, we
trace out from $\varrho$ all the other spins. This leads to \cite{wer10b}
\begin{equation}
\rho_{23}  = \left(\hspace{-.15cm}
\begin{array}{cccc}
a & 0 & 0 & 0\\
0 & b  &
c & 0 \\
0 & c &
b & 0 \\
0 & 0 & 0 & d\\
\end{array}
\hspace{-.15cm}\right), \label{rhoAB}
\end{equation}
where
\begin{eqnarray}
a &=& \frac{1+2\med{\sigma^z_2}+\med{\sigma_2^z\sigma_{3}^z}}{4}, \\
b &=& \frac{1-\med{\sigma_2^z\sigma_{3}^z}}{4}, \\
c &=& \frac{\med{\sigma_2^x\sigma_{3}^x}}{2}, \\
d &=&  \frac{1-2\med{\sigma^z_2}+\med{\sigma_2^z\sigma_{3}^z}}{4}.
\end{eqnarray}
Note that the translational symmetry of $H$ implies that $\med{\sigma^\alpha_j}=
\med{\sigma^\alpha_k}$ and $\med{\sigma^\alpha_j\sigma^\beta_{j+1}}=
\med{\sigma^\alpha_k\sigma^\beta_{k+1}}$, for any value of $j,k$.

In the thermodynamic limit, the calculation for arbitrary values of $T$, $\Delta$, 
and $h$ of the one-point correlation function 
$\med{\sigma_j^z}=Tr[\sigma_j^z\, \varrho]$ and of 
the two-point correlation functions
$\med{\sigma_j^\alpha\sigma_{j+1}^\alpha}=
Tr[\sigma_j^\alpha\sigma_{j+1}^\alpha\, \varrho]$, where $\alpha=x,z$,
was carried out in Refs. \cite{klu92,bor05,boo08,tri10} 
and reviewed in Refs. \cite{wer10b}.  In the Appendix
\ref{apA} we show the behavior of $\med{\sigma_j^z}$ and
$\med{\sigma_j^\alpha\sigma_{j+1}^\alpha}$ for several values of $T$, $\Delta$,
and $h$.

If we use Eqs.~(\ref{step0}), (\ref{stepA}), and (\ref{rhoAB}), a direct 
calculation with Eq.~(\ref{prob}) gives
\begin{eqnarray}
Q_{\Psi^{\pm}}(|\psi\rangle) &=& [1-z\cos \theta]/4, \label{qpsi}\\
Q_{\Phi^{\pm}}(|\psi\rangle) &=& [1+z\cos \theta]/4, \label{qphi}
\end{eqnarray}
where 
\begin{equation}
z = \med{\sigma^z_j}=Tr[\sigma_j^z\, \varrho]. \label{z}
\end{equation}
Contrary to the case with no field \cite{pav23}, where $Q_j(|\psi\rangle)=1/4$ 
for all $j$ and $|\psi\rangle$,
the chances of Alice measuring a given Bell state depend on the input state
$|\psi\rangle$ through $\theta$ and on the one-point correlation function $z$. 
However, averaging over the whole Bloch sphere \cite{pav23}, 
it is not difficult to see that $\langle Q_j(|\psi\rangle)\rangle=1/4$ for any $j$. Note that one should
not confuse the Bloch sphere average notation $\langle \, \rangle$ introduced
in Eq.~(\ref{Flangle}) with the 
standard notation for correlation functions as given, for instance, in Eq.~(\ref{z}).

With the aid of Eqs.~(\ref{Fidj}), (\ref{qpsi}) and (\ref{qphi}), we can compute the mean fidelity (\ref{Fbar}) for each one of the four sets of unitary operations 
available to Bob,
\begin{eqnarray}
\overline{F}(| \psi \rangle,S_{\Psi^-}) 
&=& f(r,-xx,zz), \label{f1} \\ 
\overline{F}(| \psi \rangle,S_{\Psi^+}) 
&=& f(r,xx,zz), \label{f2} \\ 
\overline{F}(| \psi \rangle,S_{\Phi^-}) 
&=& g(r,\chi,-xx,zz), \label{f3}\\
\overline{F}(| \psi \rangle,S_{\Phi^+}) 
&=& g(r,\chi,xx,zz), \label{f4}
\end{eqnarray}
where
\begin{eqnarray}
f(r,xx,zz)\!&\!\!=\!\!&\! [1 \!+ 4 r^2 (1 - r^2) (xx + zz ) \!-\! zz]/2, 
\label{f}\\
g(r,\chi,xx,zz)\!&\!\!=\!\!&\![1 + (1 - 2 r^2)^2 zz \nonumber \\
&&+ 4 r^2 (1 - r^2) xx \cos(2 \chi)]/2, \label{g}\\
xx \!&\!\!=\!\!&\! \med{\sigma_j^x\sigma_{j+1}^x} 
=Tr[\sigma_j^x\sigma_{j+1}^x\, \varrho], \label{xx}\\
zz \!&\!\!=\!\!&\! \med{\sigma_j^z\sigma_{j+1}^z}
=Tr[\sigma_j^z \sigma_{j+1}^z\, \varrho]. \label{zz}
\end{eqnarray}

Looking at Eqs.~(\ref{f}) and (\ref{g}), we realize that they do not depend on 
the one-point correlation function (\ref{z}). They only depend on the two-point
correlation functions (\ref{xx}) and (\ref{zz}). Hence, the four mean
fidelities (\ref{f1})-(\ref{f4}) depend only on the two-point correlation 
functions too. Moreover, this also implies that the expressions given by 
Eqs.~(\ref{f1})-(\ref{g}) are formally the same as the ones 
we have for the 
XXZ model without an external field \cite{pav23}. Thus, the calculations 
leading to the maximum mean fidelity and to the maximum averaged fidelity 
reported in Ref. \cite{pav23} can be literally carried over to the present case.

Maximizing over all pure states and over $S_k$ we get for the maximum mean  
fidelity \cite{pav23},
\begin{equation}
\overline{\mathcal{F}} = \max_{\{|\psi\rangle,S_k\}}{\overline{F}(|\psi\rangle,S_k)}
=\max{\left[\frac{1 + |zz|}{2}, 
\frac{1 + |xx|}{2}\right]}.
\label{fmax}
\end{equation}
The maximum or minimum of $\overline{F}(|\psi\rangle,S_k)$
occur for the input states $|\psi\rangle = |0\rangle, |1\rangle$, and 
$(|0\rangle + e^{i\chi}|1\rangle)/\sqrt{2}$. The role of these states  
in maximizing or minimizing $\overline{F}(|\psi\rangle,S_k)$ 
depends on the sign and on the magnitude of the two-point correlation functions 
$xx$ and $zz$.

On the other hand, Eq.~(\ref{Flangle}) implies that \cite{pav23}
\begin{eqnarray}
\langle \overline{F}(S_{\Psi^\pm}) \rangle &=& (3 \pm 2 xx - zz)/6, \label{fpsi+-}
\\
\langle \overline{F}(S_{\Phi^\pm}) \rangle &=& (3 + zz)/6. \label{fphi+-}
\end{eqnarray}
Maximizing over all sets $S_k$ we obtain the maximum average fidelity,
\begin{equation}
\langle\overline{\mathcal{F}}\rangle =
\max_{\{S_k\}}{\hspace{-.05cm}\langle\overline{F}(S_k)\rangle}
=\max{\hspace{-.1cm}\left[\frac{3 \!+\! 2 |xx|\!-\!zz}{6}, 
\frac{3 \!+\! zz}{6}\right]}\hspace{-.05cm}.
\label{fmed}
\end{equation}

Equations (\ref{fmax}) and (\ref{fmed}) are the two teleportation based QCP detectors
that turned out to be extremely useful and robust to detect at finite $T$ 
the QCPs for the XXZ model with no external field \cite{pav23}. 
Our goal now is to investigate their ability in detecting the QCPs for this model
when we turn on the external magnetic field.

\begin{figure}[!ht]
\includegraphics[width=8cm]{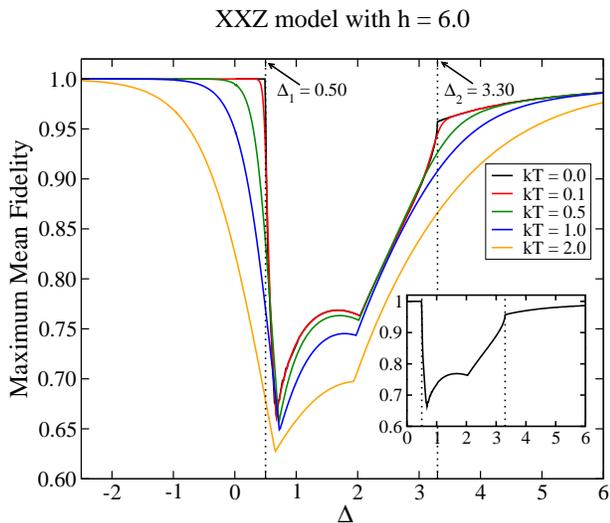}
\caption{\label{fig_max_geral6}(color online) $\overline{\mathcal{F}}$, Eq.~(\ref{fmax}), as a function of
$\Delta$ with $h=6.0$ [see Eq.~(\ref{Hxxz})]. 
At $T=0$ (see inset), both QCPs are detected by a discontinuity in
the derivatives of $\overline{\mathcal{F}}$ with respect to $\Delta$.
For $T>0$, these discontinuities in the derivatives are smoothed out. 
The maxima (or minima) of the derivatives 
are displaced away from the critical points. However, for $kT \lesssim 0.5$ these
extremum values lie close together and by extrapolating to $kT \rightarrow 0$ we 
are able to infer the correct critical points. The dotted lines mark the 
QCPs and for the solid curves the temperature increases from top to bottom. 
Here and in all other graphs all quantities are dimensionless.}
\end{figure}

\begin{figure}[!ht]
\includegraphics[width=8cm]{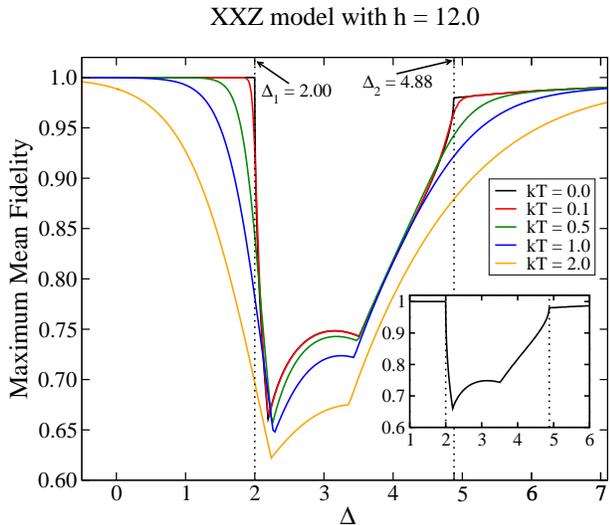}
\caption{\label{fig_max_geral12}(color online) Same as Fig. \ref{fig_max_geral6}
but now $h=12.0$. The dotted lines mark the 
QCPs and for the solid curves the temperature increases from top to bottom. }
\end{figure}

In Figs.
\ref{fig_max_geral6} and \ref{fig_max_geral12} we show $\overline{\mathcal{F}}$ 
as a function of $\Delta$ for several temperatures and for the two external fields
given in Tab. \ref{d1dinf}. 

For $T=0$ it is clear from Figs. \ref{fig_max_geral6} and \ref{fig_max_geral12},
in particular the insets, that both QCPs ($\Delta_1$ and $\Delta_{2}$)
are detected by discontinuities in the derivatives of $\overline{\mathcal{F}}$ with
respect to $\Delta$ as we cross the QCPs. 
We also see two other discontinuities in the derivatives of
$\overline{\mathcal{F}}$ for values of $\Delta$ between the two QCPs, i.e., for 
$\Delta_1 < \Delta < \Delta_{2}$. One of these extra cusp-like behavior 
for $\overline{\mathcal{F}}$ as a function of $\Delta$ is also seen when we 
study the behavior of the thermal quantum discord as a function of $\Delta$ \cite{wer10b}. 
Also, preliminary calculations \cite{pav23b} show that the
logarithm of the spectrum of the quantum coherence ($L_{QC}$) \cite{li20}
also has a cusp not related to a quantum phase transition. These cusps are robust to temperature changes since they are
not smoothed out as we increase the temperature (see Figs. \ref{fig_max_geral6} and \ref{fig_max_geral12}).

The underlying reason for these two cusps 
of $\overline{\mathcal{F}}$ is its 
particular functional form. As we change $\Delta$, 
the magnitudes of the two-point correlation functions $xx$ and $zz$ change. 
As we cross the two cusps, the correlation function with the greater magnitude changes. 
This change is reflected in a discontinuity in the value of $\overline{\mathcal{F}}$
[see Eq.~(\ref{fmax})].

\begin{figure}[!ht]
\includegraphics[width=8cm]{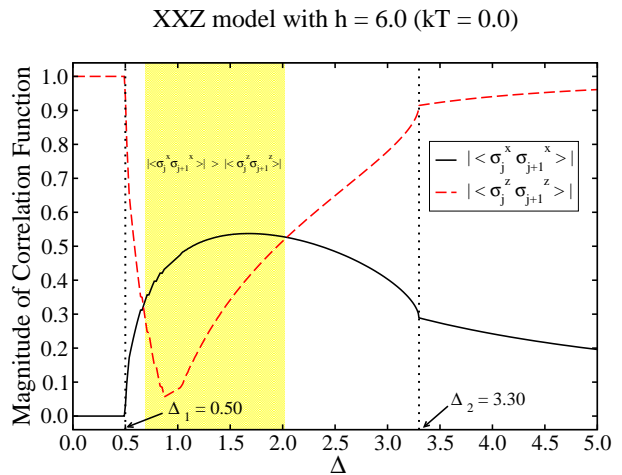}
\caption{\label{fig_mod}(color online) Magnitudes of 
$\med{\sigma_j^x\sigma_{j+1}^x}$ and $\med{\sigma_j^z\sigma_{j+1}^z}$
as a function of $\Delta$ when $h=6.0$ and $T=0$. The yellow-shaded region
is the region between the two cusps of $\overline{\mathcal{F}}$ that do not
correspond to QCPs.}
\end{figure}

As an illustrative example, in Fig. \ref{fig_mod} we show for $T=0$ the behavior of 
$|xx|$ and $|zz|$ as a function of $\Delta$ assuming $h=6.0$. A similar behavior is seen for $T>0$ and also when we have $h=12.0$. Looking at Fig. \ref{fig_mod}, it is clear that $|xx|>|zz|$ in the yellow-shaded region, while $|xx|<|zz|$ outside that region. The yellow-shaded region was drawn such that it represents 
the region between the two cusps of $\overline{\mathcal{F}}$ 
that are not related to QPTs (see Fig. \ref{fig_max_geral6}). Looking at 
Fig. \ref{fig_mod}, it is clear that the boundaries of the yellow-shaded region
coincide with the two values of $\Delta$ in which the roles of $xx$ and $zz$ are 
exchanged in the evaluation of (\ref{fmax}) and (\ref{fmed}).
It is this property 
of $|xx|$ and $|zz|$ as we cross those two points that causes the two extra 
cusps seen in $\overline{\mathcal{F}}$.

It is worth mentioning that when we do not have an external field ($h=0$), 
the QCPs $\Delta_1$ and $\Delta_{2}$ are located exactly at the points
in which $|xx|=|zz|$. This is why $\overline{\mathcal{F}}$ is very robust in 
detecting those two QCPs for finite $T$, retaining its cusp-like behavior at 
the QCPs as we increase $T$ \cite{pav23}. When $h=0$ we only have two discontinuities,
exactly at the locations of the two QCPs \cite{pav23}. 
When $h\neq 0$, 
on the other hand, the points where $|xx|=|zz|$
are shifted away from the QCPs
and four cusps instead of two 
are seen when $T=0$. 
Two of them are related to the two QCPs and the other two are associated with
the points where $|xx|=|zz|$. 

\begin{figure}[!ht]
\includegraphics[width=8cm]{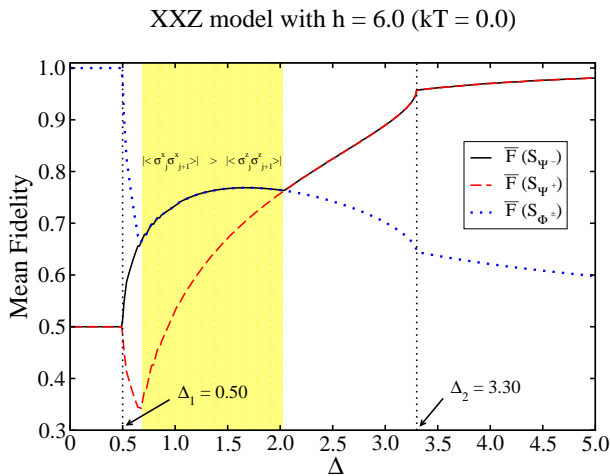}
\caption{\label{fig_fid1a4}(color online) $\overline{F}(S_{k})$, Eq.~(\ref{fsk}), 
as a function of $\Delta$ when $T=0$ and $h=6.0$.}
\end{figure}

We can also better understand the behavior of $\overline{\mathcal{F}}$ if we analyze
the behavior of the following quantity, 
\begin{equation}
\overline{F}(S_k)= \max_{\{|\psi\rangle\}}{\overline{F}(|\psi\rangle,S_k)}. 
\label{fsk}
\end{equation}
Equation (\ref{fsk}) is obtained from $\overline{F}(|\psi\rangle,S_k)$
by maximizing it over all input states only. In this way, as we show in
Fig. \ref{fig_fid1a4}, we are able to study
how $\overline{F}(S_k)$ behaves for each one of the possible values of $k$,
\begin{eqnarray}
\overline{F}(S_{\Psi^\pm}) &=& \max{\left[\frac{1 - zz}{2}, \frac{1 \pm xx}{2}\right]},
\label{f1max}\\
\overline{F}(S_{\Phi^\pm}) &=& \max{\left[\frac{1 + zz}{2}, \frac{1 + |xx|}{2}\right]}.
\label{f4max}
\end{eqnarray}

Looking at Fig. \ref{fig_fid1a4}, we notice that 
before $\Delta_1$ (the first QCP) and up to where $|xx|=|zz|$ (before the 
yellow-shaded region), the maximum mean fidelity  
$\overline{\mathcal{F}}$ is given by 
$\overline{F}(S_{\Phi^\pm})$. In the yellow-shaded region, 
we have either $\overline{F}(S_{\Phi^\pm})$ or 
$\overline{F}(S_{\Psi^-})$ as the maximum mean fidelity. After the yellow-shade
region, $\overline{F}(S_{\Psi^\pm})$ dominates.
Furthermore, the points where the roles of $\overline{F}(S_{\Psi^\pm})$
and $\overline{F}(S_{\Phi^\pm})$ are exchanged in furnishing the greatest
mean fidelity occur exactly where $|xx|=|zz|$ (the boundaries of the 
yellow-shaded region). This is why we see the two cusps  
of $\overline{\mathcal{F}}$ that are not related to QPTs. The other two
derivative discontinuities, 
associated with the two QCPs, are due to the particular behavior
of the two-point correlation functions at those points. The discontinuities in the
derivatives of $zz$ in the first and second QCPs are reflected in the discontinuities
of the derivatives of $\overline{\mathcal{F}}$ at those points
(cf. Figs. \ref{fig_mod} and \ref{fig_fid1a4}). Had we worked 
with the minimum mean fidelity \cite{pav23}, the relevant two-point correlation function would be $xx$. 

When $T>0$, the cusps located at the two QCPs are smoothed out and
displaced away. The 
other two cusps are not smoothed out although being displaced too. 
As such, in order to determine the two QCPs in this
scenario, we follow a similar strategy used in Ref. \cite{wer10b} to deal with
the smoothing out of the cusps of the thermal quantum discord around the QCPs at 
finite $T$. As we increase the
temperature, the discontinuities in the derivatives of $\overline{\mathcal{F}}$
that occur exactly at the QCPs when $T=0$ are now manifested in very high values 
for the magnitudes of those derivatives, with those maxima displaced from the correct locations of the QCPs.
However, for $kT \lesssim 0.5$
the maximum (or minimum) of the derivatives as a function of $kT$
lie more or less along a straight line 
and by extrapolating to zero from a few finite $T$ data we can correctly predict the exact locations of the two QCPs. 

\begin{figure}[!ht]
\includegraphics[width=8cm]{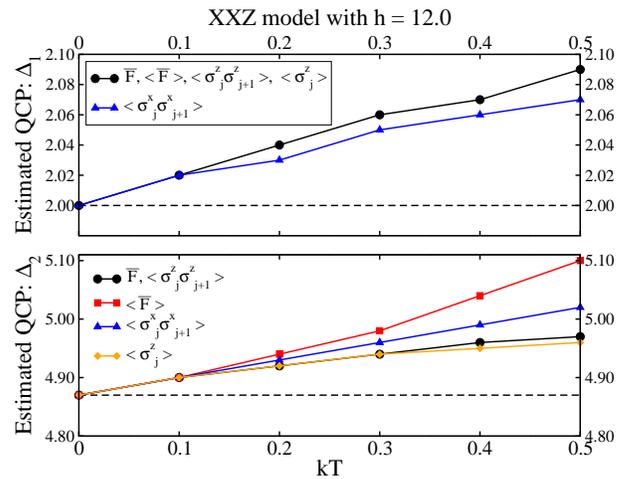}
\caption{\label{fig_qcpT} (color online) Estimated QCPs  
after determining the extrema of the first order (upper panel) and 
of the second order (lower panel) derivatives with respect to $\Delta$ 
for $\overline{\mathcal{F}}, \langle\overline{\mathcal{F}}\rangle, 
\langle \sigma^x_j\sigma^x_{j+1} \rangle, \langle \sigma^z_j\sigma^z_{j+1} \rangle$, and $\langle \sigma^z_j\rangle$ at several different values of $T$. 
See text for details on how the 
QCPs were estimated. The dashed lines mark the exact values of the QCPs.}
\end{figure}

In the upper panel of Fig. \ref{fig_qcpT}, we show as a function of $T$  
the values of $\Delta$ where we find the maximum of $|dy/d\Delta|$, with $y$ representing the quantities shown in Fig. \ref{fig_qcpT}.
In the upper panel we picked the maxima of $|dy/d\Delta|$ around $\Delta_1$. 
In the lower panel of Fig. \ref{fig_qcpT}, we show as a function of $T$ the 
spots of the maximum values of $|d^2y/d\Delta^2|$ about $\Delta_{2}$. 
Note that although in Fig. \ref{fig_qcpT} 
we chose the external field to be $h=12.0$, the analysis reported below applies 
to other values of fields as well. 

For $kT=0, 0.1, 0.2, 0.3, 0.4$, and $0.5$,
we computed $\overline{\mathcal{F}}$, $\langle\overline{\mathcal{F}}\rangle$, and 
the one- and two-point correlation functions as a function of $\Delta$ in increments
of $0.01$. Then, we numerically computed the first order derivatives of 
those quantities about $\Delta_1$ and their second order derivatives about 
$\Delta_{2}$. The values of $\Delta$ leading to the
greatest values for the magnitudes of those derivatives are shown in 
Fig. \ref{fig_qcpT}. If we take into account that $\Delta$ was changed in 
increments of $0.01$, the spots of the 
maxima of the magnitudes of the first order derivatives are obtained 
within an accuracy of $\pm 0.01$ about the values shown in the upper panel of 
Fig. \ref{fig_qcpT}. And since the second order derivatives are obtained
from the first order ones, which already have a numerical error of $0.01$, we estimate 
that the error for the location of the maxima of the absolute values of the 
second order derivatives are at least $\pm 0.02$ about the values shown in the lower panel. Excluding the data for
$T=0$, we made linear regressions with the remaining data 
($kT=0.1, 0.2, 0.3, 0.4, 0.5$) in order to check whether a straight line would 
correctly predict the QCPs at $T=0$. For all quantities shown in the upper panel of Fig. \ref{fig_qcpT} and to all but one in the lower panel, the linear coefficients
(y-axis intercepts)
correctly predicted the QCPs within an accuracy of $\pm 0.01$. For 
$\langle\overline{\mathcal{F}}\rangle$, however, we needed a quadratic regression 
to extrapolate to the correct value of $\Delta_{2}$ with an accuracy of 
$\pm 0.01$.

\begin{figure}[!ht]
\includegraphics[width=8cm]{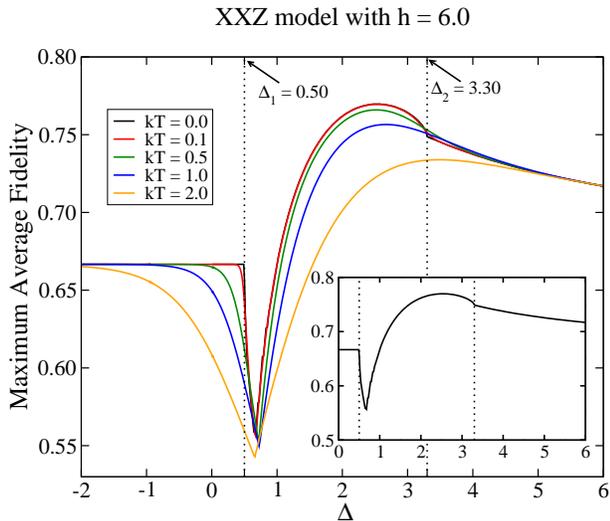}
\caption{\label{fig_med_geral6}(color online) $\langle\overline{\mathcal{F}}\rangle$, Eq.~(\ref{fmed}), as a function of
$\Delta$ with $h=6.0$ [see Eq.~(\ref{Hxxz})]. 
At $T=0$ (see inset), both QCPs are detected by a discontinuity in
the derivatives of $\langle\overline{\mathcal{F}}\rangle$ with respect to $\Delta$.
For $T>0$, these discontinuities in the derivatives are smoothed out. 
The maximum (or minimum) of the derivatives 
are displaced away from the critical points. However, for $kT \lesssim 0.5$ these
extremum values lie close together and by extrapolating to $kT \rightarrow 0$ we 
are able to infer the correct critical points. The dotted lines mark the 
QCPs and for the solid curves the temperature increases from top to bottom.}
\end{figure}

\begin{figure}[!ht]
\includegraphics[width=8cm]{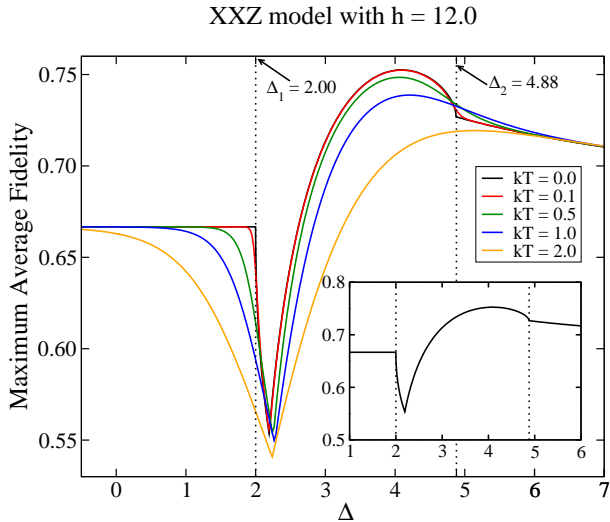}
\caption{\label{fig_med_geral12}(color online) Same as Fig. \ref{fig_med_geral6}
but now $h=12.0$. The dotted lines mark the 
QCPs and for the solid curves the temperature increases from top to bottom.}
\end{figure}

To end this section we show in Figs. \ref{fig_med_geral6} and \ref{fig_med_geral12}
the behavior of $\langle\overline{\mathcal{F}}\rangle$ as a function of $\Delta$
for several temperatures and for the two external fields shown in Tab. \ref{d1dinf}.

\begin{figure}[!ht]
\includegraphics[width=8cm]{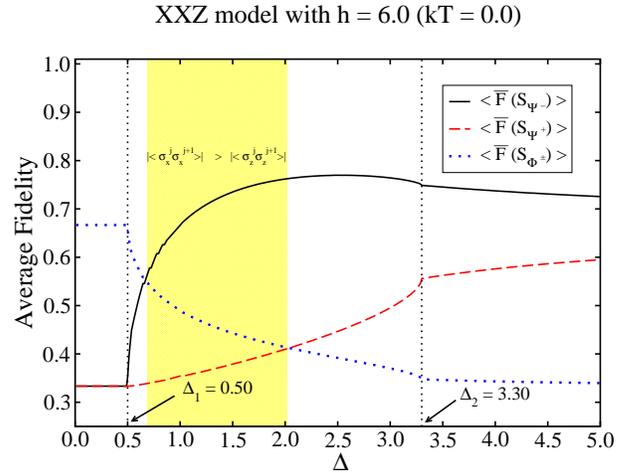}
\caption{\label{fig_fidmed1a4}(color online) $\langle\overline{F}(S_{k})\rangle$, Eqs.~(\ref{fpsi+-}) and (\ref{fphi+-}), 
as a function of $\Delta$ when $T=0$ and $h=6.0$.}
\end{figure}

Looking at Figs.~\ref{fig_med_geral6} and \ref{fig_med_geral12}, we realize that now,
contrary to the behavior of  $\overline{\mathcal{F}}$ 
(see Figs. \ref{fig_max_geral6} and \ref{fig_max_geral12}), 
$\langle\overline{\mathcal{F}}\rangle$ has only three instead of four derivative 
discontinuities at $T=0$. Two of them are related to the two QCPs for this 
model while the remaining one is associated with the particular functional form 
of $\langle\overline{\mathcal{F}}\rangle$ [see Eq.~(\ref{fmed})]. This third cusp 
of $\langle\overline{\mathcal{F}}\rangle$ is located at one of the values of 
$\Delta$ in which $|xx|=|zz|$ (left boundary of the yellow-shaded region in 
Fig. \ref{fig_fidmed1a4}). 

In order to understand the absence of the fourth cusp for 
$\langle \overline{\mathcal{F}} \rangle$ at $T=0$, we study the individual 
behavior of  $\langle \overline{F}(S_{\Psi^\pm}) \rangle$ and 
$\langle \overline{F}(S_{\Phi^\pm}) \rangle$,
Eqs.~(\ref{fpsi+-}) and (\ref{fphi+-}), as a function of $\Delta$.
Since $\langle \overline{\mathcal{F}} \rangle$ is obtained by picking the
greatest value among these four quantities, by tracing back which quantity 
gives $\langle \overline{\mathcal{F}} \rangle$ we can understand the origin of the cusp-like behavior of $\langle \overline{\mathcal{F}} \rangle$.   

In Fig. \ref{fig_fidmed1a4} we show 
$\langle \overline{F}(S_{k}) \rangle$ for $k=\Psi^\pm,\Phi^\pm$ and fixing $h=6.0$ 
(a similar plot applies to $h=12.0$). 
Fig. \ref{fig_fidmed1a4} tells us that before the first QCP $\Delta_1$ and up to
where $|xx|=|zz|$ for the first time, the maximum average fidelity  
$\langle\overline{\mathcal{F}}\rangle$ is given by 
$\langle\overline{F}(S_{\Phi^\pm})\rangle$. Inside the yellow-shaded region,
where $|xx| > |zz|$, and way up to and beyond the second QCP $\Delta_{2}$,  
the value of $\langle\overline{\mathcal{F}}\rangle$ is dictated by 
$\langle\overline{F}(S_{\Psi^-})\rangle$. There is no change of the function 
that maximizes $\langle\overline{\mathcal{F}}\rangle$ at the right boundary 
of the yellow-shaded region, contrary to what we see for 
$\overline{\mathcal{F}}$ (Fig. \ref{fig_fid1a4}). That is the reason we do not have a cusp where the two-point correlation functions become equal again ($|xx|=|zz|$),
at the right boundary of the yellow-shaded region. Furthermore, the two cusps
related to the QCPs have their origin in the intrinsic functional form
of $\langle\overline{\mathcal{F}}\rangle$ that is not associated with 
$\langle\overline{F}(S_{\Psi^\pm})\rangle$ and $\langle\overline{F}(S_{\Phi^\pm})\rangle$ changing their roles in maximizing 
$\langle\overline{\mathcal{F}}\rangle$. Indeed, the cusps of 
$\langle\overline{\mathcal{F}}\rangle$ at the two QCPs are a consequence of the 
cusps observed for the two-point correlation functions at those points. Since 
$\langle\overline{\mathcal{F}}\rangle$ is a linear function of those correlation
functions, any discontinuities in their derivatives with respect to $\Delta$ will
manifest themselves in discontinuities of the derivatives of 
$\langle\overline{\mathcal{F}}\rangle$ (see Appendix \ref{apA}).

For $T>0$, and similarly to the case of $\overline{\mathcal{F}}$, 
the cusps at the two QCPs that we see for 
$\langle\overline{\mathcal{F}}\rangle$ at $T=0$
are smoothed out and
displaced away. The other remaining cusp is not smoothed out although being 
displaced too. The procedure to estimate the QCPs using finite $T$ data in the 
present case is exactly the same one reported for $\overline{\mathcal{F}}$ a 
few paragraphs ago and the results of this analysis are given in Fig. \ref{fig_qcpT}.

\section{The XY and the Ising model}
\label{secXY}

Using the notation of Sec. \ref{secXXZ}, 
the anisotropic one-dimensional XY model subjected to a transverse magnetic field is 
described by the following Hamiltonian \cite{lie61,bar70,bar71},
\begin{equation}
H \!=\! -\frac{\lambda}{4}\!\sum_{j=1}^{L}\!\left[(1+\gamma)\sigma^{x}_{j}\sigma^{x}_{j+1} + (1-\gamma)\sigma^{y}_{j}\sigma^{y}_{j+1}\right]\! 
- \frac{1}{2}\!\sum_{j=1}^{L}\!\sigma^z_j, \label{Hxy}
\end{equation}
with $\lambda$ being related to the inverse of the  
external magnetic field strength and
$\gamma$ the anisotropy parameter. If we set $\gamma=\pm 1$ we have the 
transverse Ising model and for $\gamma=0$ we obtain the XX model in a  transverse
field.

As we change $\lambda$ (essentially the external field), 
the ground state for the XY model goes through a  
QPT when we reach the QCP $\lambda_c=1.0$. This is 
the Ising transition, where for $\lambda < 1$ we have a ferromagnetic ordered
phase and for $\lambda > 1$ we have a quantum paramagnetic phase \cite{pfe70}.
Whenever $\lambda > 1$, we also observe another 
QPT as we change 
the anisotropy parameter $\gamma$. 
It is called the anisotropy transition and it occurs at $\gamma_c=0$ 
\cite{lie61,bar70,bar71,zho10}. This QPT separates a ferromagnet
ordered in the x-direction from a ferromagnet ordered in the y-direction.
Although the $\lambda$ and $\gamma$ QPTs above are of the same order,
they belong to different universality classes \cite{lie61,bar70,bar71,zho10}.

The canonical ensemble density matrix describing the whole spin chain in
equilibrium with a heat bath of temperature $T$ is $\varrho=e^{-H/kT}/Z$ and 
the density matrix describing a pair of nearest neighbors, obtained after
tracing out all but those two spins, is \cite{osb02,wer10b}
\begin{equation}
\rho_{23}  = \left(\hspace{-.15cm}
\begin{array}{cccc}
a & 0 & 0 & e\\
0 & b  & c & 0 \\
0 & c & b & 0 \\
e & 0 & 0 & d\\
\end{array}
\hspace{-.15cm}\right), \label{rhoAB2}
\end{equation}
where
\begin{eqnarray}
a &=& \frac{1+2\med{\sigma^z_2}+\med{\sigma_2^z\sigma_{3}^z}}{4}, \\
b &=& \frac{1-\med{\sigma_2^z\sigma_{3}^z}}{4}, \\
c &=& \frac{\med{\sigma_2^x\sigma_{3}^x}+\med{\sigma_2^y\sigma_{3}^y}}{4}, \\
d &=&  \frac{1-2\med{\sigma^z_2}+\med{\sigma_2^z\sigma_{3}^z}}{4}, \\
e &=& \frac{\med{\sigma_2^x\sigma_{3}^x}-\med{\sigma_2^y\sigma_{3}^y}}{4}.
\end{eqnarray}
Similarly to the XXZ model of Sec. \ref{secXXZ}, the translational symmetry of 
the XY model implies that $\med{\sigma^\alpha_j}=
\med{\sigma^\alpha_k}$ and $\med{\sigma^\alpha_j\sigma^\beta_{j+1}}=
\med{\sigma^\alpha_k\sigma^\beta_{k+1}}$, for any value of $j,k$.

The computation in the thermodynamic limit and 
for arbitrary values of $T$, $\lambda$, and $\gamma$ 
of the one-point correlation function 
$z=\med{\sigma_j^z}=Tr[\sigma_j^z\, \varrho]$ and of 
the two-point correlation functions
$\alpha\alpha=\med{\sigma_j^\alpha\sigma_{j+1}^\alpha}=
Tr[\sigma_j^\alpha\sigma_{j+1}^\alpha\, \varrho]$, where $\alpha=x,y,z$,
is given in Refs. \cite{lie61,bar70,bar71}. In Ref. \cite{wer10b} this solution is
written in the present notation and in the Appendix
\ref{apB} we show the behavior of $\med{\sigma_j^z}$ and
$\med{\sigma_j^\alpha\sigma_{j+1}^\alpha}$ for $T=0$ and several values of 
$\lambda$ and $\gamma$. We also give
a brief qualitative discussion of how they differ from the $T=0$ case. 

Proceeding 
along the same lines as
in Sec. \ref{secXXZ}, inserting
Eqs.~(\ref{step0}), (\ref{stepA}), and (\ref{rhoAB2}) into Eq.~(\ref{prob}) 
leads to the same set of probabilities $Q_j(|\psi\rangle)$ for Alice measuring
a given Bell state [cf. Eqs.~(\ref{qpsi}) and (\ref{qphi})].
Using Eqs.~(\ref{Fidj}), (\ref{qpsi}) and (\ref{qphi}), the mean fidelity (\ref{Fbar}) for each one of Bob's four sets of unitary operations 
become
%
%
%
%
\begin{eqnarray}
\overline{F}(| \psi \rangle,S_{\Psi^-}) 
&=& h(r,\chi,-xx,-yy,zz), \label{h1} \\ 
\overline{F}(| \psi \rangle,S_{\Psi^+}) 
&=& h(r,\chi,xx,yy,zz), \label{h2} \\ 
\overline{F}(| \psi \rangle,S_{\Phi^-}) 
&=& h(r,\chi,-xx,yy,-zz), \label{h3}\\
\overline{F}(| \psi \rangle,S_{\Phi^+}) 
&=& h(r,\chi,xx,-yy,-zz), \label{h4}
\end{eqnarray}
where
\begin{eqnarray}
h(r,\chi,xx,yy,zz)\hspace{-.15cm}&=&\hspace{-.15cm} [1 + 2 r^2 (1 - r^2) (xx + yy + 2 zz) -zz\nonumber \\
\hspace{-.15cm}&+&\hspace{-.15cm} 2 r^2(1-r^2)(xx-yy)\cos(2\chi)]/2. 
\label{h}
\end{eqnarray}
Note that if we assume $xx=yy$ in Eqs.~(\ref{h1})-(\ref{h4}), we obtain 
the corresponding expressions for the XXZ model, namely, 
Eqs.~(\ref{f1})-(\ref{f4}). 
That this should indeed occur can be seen by setting
$xx=yy$ in the two-qubit density matrix (\ref{rhoAB2}). In this case we 
recover the two-qubit density matrix for the XXZ model, Eq.~(\ref{rhoAB}),
and consequently Eqs.~(\ref{f1})-(\ref{f4}) must follow from Eqs.~(\ref{h1})-(\ref{h4}) if we assume $xx=yy$.

Repeating the calculations of Ref. \cite{pav23} that led to the optimum 
mean fidelity over all input states, it is not difficult
to see that the extrema of Eq.~(\ref{Fbar}) occur for the states
$|\psi\rangle = |0\rangle, |1\rangle$, and 
$(|0\rangle + e^{i\chi}|1\rangle)/\sqrt{2}$. This gives
\begin{eqnarray}
\overline{F}(S_{\Psi^\pm}) &=& \max{\left[\frac{1 \pm xx}{2}, \frac{1 \pm yy}{2},
\frac{1 - zz}{2}\right]},
\label{h1max}\\
\overline{F}(S_{\Phi^\pm}) &=& \max{\left[\frac{1 \pm xx}{2}, \frac{1 \mp yy}{2},
\frac{1 + zz}{2}\right]},
\label{h4max}
\end{eqnarray}
where
$\overline{F}(S_k)$ is given by Eq.~(\ref{fsk}). If we now maximize over the four possible sets of unitary operations available to Bob, we get the maximum 
of the mean fidelity (\ref{Fbar}) for the present model,
\begin{eqnarray}
\overline{\mathcal{F}} &=& \max_{\{|\psi\rangle,S_k\}}{\overline{F}(|\psi\rangle,S_k)} \nonumber \\ 
&&=\max{\left[\frac{1 + |xx|}{2}, \frac{1 + |yy|}{2}, 
\frac{1 + |zz|}{2}\right]}.
\label{fmax2}
\end{eqnarray}

Averaging over all input states lying on the Bloch sphere \cite{pav23}, we get
from Eqs.~(\ref{Flangle}) and (\ref{h1})-(\ref{h4}),
\begin{eqnarray}
\langle \overline{F}(S_{\Psi^\pm}) \rangle &=& (3 \pm xx \pm yy - zz)/6, \label{hpsi+-}
\\
\langle \overline{F}(S_{\Phi^\pm}) \rangle &=& (3 \pm xx \mp yy + zz)/6. \label{hphi+-}
\end{eqnarray}
Using Eqs.~(\ref{hpsi+-}) and (\ref{hphi+-}), the maximum average fidelity is 
\begin{eqnarray}
\hspace{-4cm}\langle\overline{\mathcal{F}}\rangle \hspace{-.15cm}&=&\hspace{-.15cm}
\max_{\{S_k\}}{\hspace{-.05cm}\langle\overline{F}(S_k)\rangle} \nonumber \\
\hspace{-.15cm}&=&\hspace{-.15cm}
\max\hspace{-.1cm}
{\left[\frac{3 + |xx+yy|-zz}{6}, \frac{3 +|xx-yy| + zz}{6}\right]}\hspace{-.05cm}.
\label{fmed2}
\end{eqnarray}

Equations (\ref{fmax2}) and (\ref{fmed2}) are the analogs of Eqs. (\ref{fmax}) 
and (\ref{fmed}) for the present model. Note that if $xx=yy$, we recover
Eqs.~(\ref{fmax}) and (\ref{fmed}) from (\ref{fmax2}) and (\ref{fmed2}). 

We now focus on studying the efficiency of Eqs.~(\ref{fmax2}) and 
(\ref{fmed2}) in detecting the QCPs for the XY model in a transverse field at
zero and non-zero temperatures. Following Ref. \cite{wer10b}, we expect that 
the best way to pinpoint the QCP $\lambda_c$ for the XY model is by studying the first and second order derivatives of Eqs.~(\ref{fmax2}) and (\ref{fmed2}) with respect to $\lambda$.
It turns out that 
the extremum values of the derivatives are located at this QCP for $T=0$ and 
move away as we increase $T$. For a sufficiently low range of temperatures, 
these extremum values lie on a straight line and by extrapolating to $T=0$ we can predict the correct QCP. Also, our numerical analysis showed that 
$\overline{\mathcal{F}}$ is a better QCP detector than 
$\langle\overline{\mathcal{F}}\rangle$ when it comes to spotlighting the 
Ising transition ($\lambda_c$), with the former having greater and 
sharper maximum (or minimum) about this QCP. Therefore, here we only show the 
behavior of $\overline{\mathcal{F}}$ about this QCP. 

\begin{figure}[!ht]
\includegraphics[width=8cm]{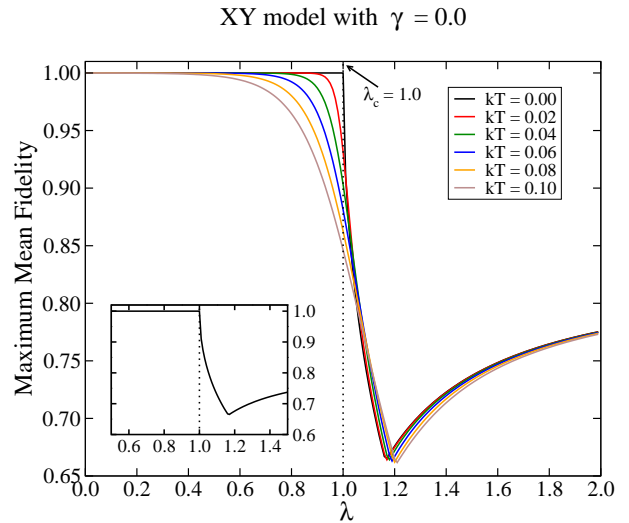}
\caption{\label{fig_max_geral_xy}(color online) $\overline{\mathcal{F}}$, Eq.~(\ref{fmax2}), as a function of
$\lambda$ with $\gamma=0.0$ (XX model in a transverse field) [see Eq.~(\ref{Hxy})]. 
At $T=0$ (see the inset),  the QCP $\lambda_c=1.0$ is detected by a discontinuity in
the derivatives of $\overline{\mathcal{F}}$ with respect to $\lambda$.
For $T>0$, these discontinuities in the derivatives are smoothed out. 
The maximum (or minimum) of the derivatives 
moves away from the QCP. However, for $kT \lesssim 0.1$ these
extremum values lie in a straight line and by extrapolating to $kT \rightarrow 0$ 
we can discover the right value for the QCP. The dotted lines mark the 
QCP $\lambda_c$ and for the solid curves the temperature increases from top to bottom when $\lambda < \lambda_c$.}
\end{figure}

\begin{figure}[!ht]
\includegraphics[width=8cm]{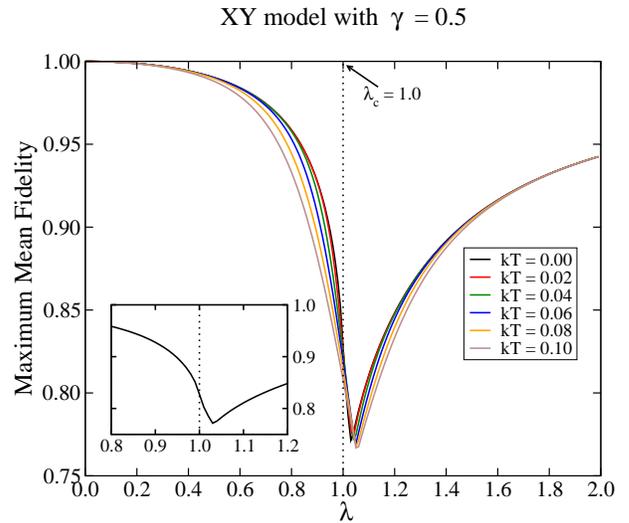}
\caption{\label{fig_max_geral_xy2}(color online) Same as Fig. \ref{fig_max_geral_xy}
but now $\gamma=0.5$. The dotted lines mark the 
QCP $\lambda_c$ and for the solid curves the temperature increases from top to bottom when $\lambda < \lambda_c$.}
\end{figure}

\begin{figure}[!ht]
\includegraphics[width=8cm]{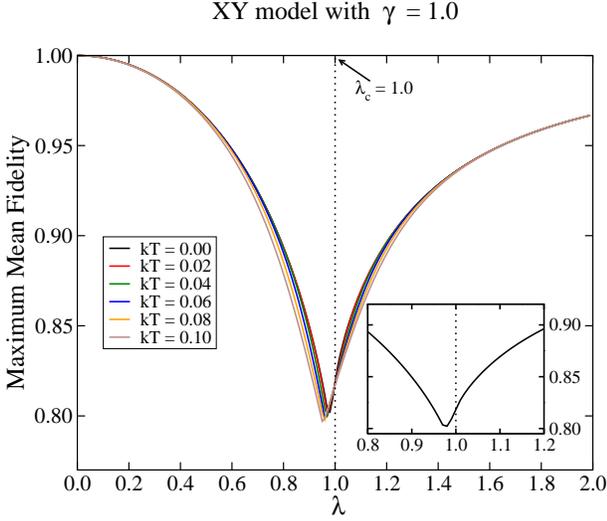}
\caption{\label{fig_max_geral_xy3}(color online) Same as Fig. \ref{fig_max_geral_xy}
but now $\gamma=1.0$ (Ising model in a transverse field).
The dotted lines mark the 
QCP $\lambda_c$ and for the solid curves the temperature increases from top to bottom before the kinks (minima).}
\end{figure}

In Figs. \ref{fig_max_geral_xy}, \ref{fig_max_geral_xy2}, and
\ref{fig_max_geral_xy3} we show for several values of $T$ and $\gamma$ 
the behavior of $\overline{\mathcal{F}}$ as a function of $\lambda$. 
For $\gamma = 0.0$ we have the isotropic XX model in a transverse field. 
Looking at Fig. \ref{fig_max_geral_xy}, we realize that at $T=0$ the QCP is 
determined by a discontinuity in the first derivative of $\overline{\mathcal{F}}$. 
For high values of $T$ this discontinuity in the derivative is smoothed out 
and displaced from the exact location of the QCP, namely, $\lambda_c=1.0$.
However, as we will show in a moment, we still can determine the correct QCP
using finite $T$ data.

For the other values of $\gamma$, i.e., $\gamma=0.5$ (anisotropic XY model in 
a transverse field) and $\gamma=1.0$ (Ising model in a transverse field), the
QCP is determined by an inflection point that occurs exactly at $\lambda_c=1.0$
when $T=0$. As we increase $T$, this inflection point moves away 
from $\lambda_c$ and by determining the maximum (minimum) of the first and second 
order derivatives of $\overline{\mathcal{F}}$ with respect to $\lambda$ we can infer the correct QCP extrapolating from finite $T$ data. 

Two remarks are in order now. First, 
the behavior of $\overline{\mathcal{F}}$ for all $T$ and
$\gamma$ around $\lambda_c$ is similar to the behavior of the two-point 
correlation functions about that point. This is true because  
$\overline{\mathcal{F}}$ is essentially a linear function of the two-point 
correlation functions in the neighborhood of the QCP [cf. Eq.~(\ref{fmax2})].
Being more specific, the derivatives of $\overline{\mathcal{F}}$ at and 
about the QCP are proportional to the derivatives of the two-point correlation function furnishing the greatest magnitude at and in the neighborhood of the QCP.
Therefore, the functional behavior of $\overline{\mathcal{F}}$ and its derivatives about the QCP is essentially the same as this correlation function about the QCP.
Second, the cusp-like behavior seen in Figs. \ref{fig_max_geral_xy},
\ref{fig_max_geral_xy2}, and \ref{fig_max_geral_xy3} slightly away from 
the QCP is related to the point where $|xx|=|zz|$ (see Figs. \ref{fig1_apB}-\ref{fig3_apB} in the Appendix \ref{apB}). Before the cusp, $|zz|>|xx|$,
and after it, $|zz|<|xx|$. It is this fact and the particular functional 
form of $\overline{\mathcal{F}}$ that lead to those cusps. This is similar to
what we have found when dealing with the XXZ model in Sec. \ref{secXXZ}.

\begin{figure}[!ht]
\includegraphics[width=8cm]{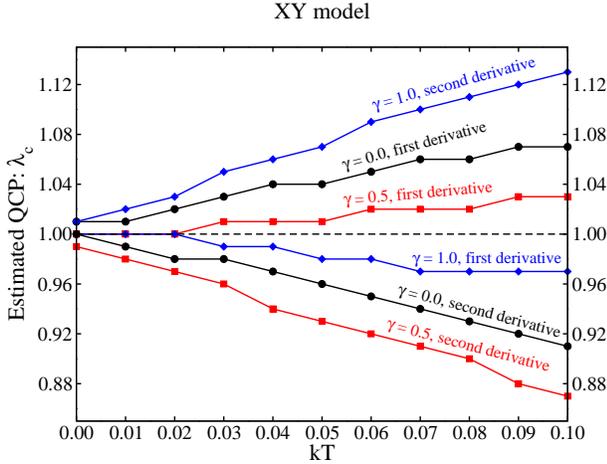}
\caption{\label{fig_qcpTxy} (color online) Estimated value for the QCP  
$\lambda_c$ using the location of the maximum (or minimum) of the first and 
second order derivatives of $\overline{\mathcal{F}}$ with respect to $\lambda$
for several values of $T$. See text for details. 
The dashed line gives the exact value of the QCP.}
\end{figure}

Returning to the analysis of how to obtain the correct QCP using finite $T$
data, we follow Ref. \cite{wer10b} and the procedure already explained in Sec. \ref{secXXZ},
i.e., we numerically compute the first and second order derivatives of 
$\overline{\mathcal{F}}$ around the QCP and search for their extremum values as indicators of a QPT.
In Fig. \ref{fig_qcpTxy} we plot as a function of $T$ the value of $\lambda$
(y-axis) furnishing the extrema of the first and second 
order derivatives of $\overline{\mathcal{F}}$ with respect to $\lambda$
in the neighborhood of the QCP $\lambda_c$. 

For the eleven values of $kT$ shown in Fig. \ref{fig_qcpTxy}, namely, 
$kT = 0.00, 0.01, 0.02, \ldots, 0.10$, we computed 
$\overline{\mathcal{F}}$ as a function of $\lambda$ in increments of $0.01$. 
Subsequently, we numerically obtained its first and second order derivatives with
respect to $\lambda$. The points shown in Fig. \ref{fig_qcpTxy} are the location
of the extrema of those derivatives. Similarly to what we had for the XXZ model,
the locations of those extrema are obtained within an accuracy of $\pm 0.01$ 
for the first derivatives and $\pm 0.02$ for the second derivatives. 

Dropping the data for $kT = 0.00$, we implemented a simple linear regression  
with the remaining data ($kT = 0.01, 0.02, \ldots, 0.10$) to verify if 
a straight line could correctly predict the exact location of the 
QCP at $kT = 0$. For the six curves shown Fig. \ref{fig_qcpTxy}, 
the obtained linear coefficients (y-axis intercepts) predicted with 
an accuracy of $0.01$ the correct location of $\lambda_c$. 

\begin{figure}[!ht]
\includegraphics[width=8cm]{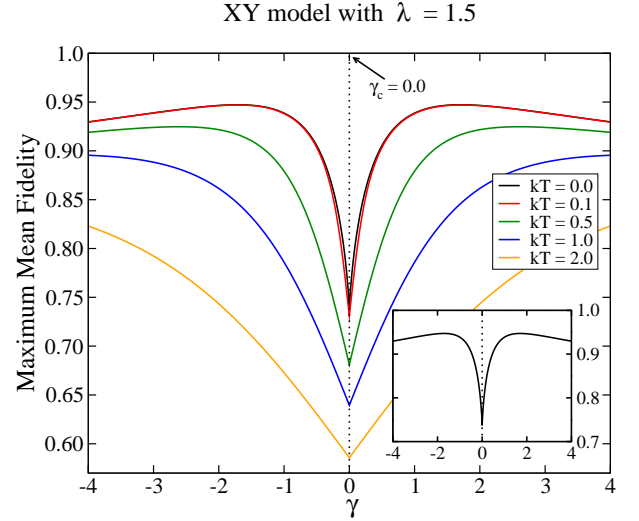}
\caption{\label{fig_max_geral_gamma}(color online) 
$\overline{\mathcal{F}}$, Eq.~(\ref{fmax2}), as a function of
$\gamma$ with $\lambda=1.5$ [see Eq.~(\ref{Hxy})]. 
Both at $T=0$ (see inset) and $T>0$, the QCP $\gamma_c=0.0$ is detected by a
cusp that occurs exactly at the location of the QPT. The dotted lines represent the 
QCP $\gamma_c$ and for the solid curves the temperature increases from top 
to bottom.}
\end{figure}

\begin{figure}[!ht]
\includegraphics[width=8cm]{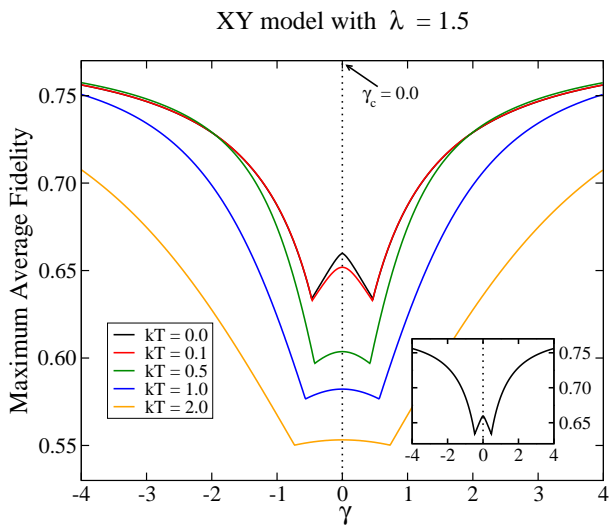}
\caption{\label{fig_med_geral_gamma}(color online) 
Same as Fig. \ref{fig_max_geral_gamma} but now we have 
$\langle\overline{\mathcal{F}}\rangle$, Eq.~(\ref{fmed2}),
as a function of $\gamma$. In this case the QCP is detected by a cusp at $T=0$,
with the latter being smoothed out as we increase $kT$.
The dotted lines represent the 
QCP $\gamma_c$ and for the solid curves the temperature increases from top 
to bottom.}
\end{figure}

Finally, in Figs. \ref{fig_max_geral_gamma} and \ref{fig_med_geral_gamma}
we show, respectively, $\overline{\mathcal{F}}$ and 
$\langle\overline{\mathcal{F}}\rangle$ as functions of $\gamma$, fixing 
$\lambda=1.5$. It is clear from the plots in both figures that 
the anisotropy QPT is clearly detected by both the maximum mean and 
maximum average fidelities. 

The QCP $\gamma_c=0.0$ is detected by a cusp-like behavior of 
$\overline{\mathcal{F}}$ for all values of $T$ shown in 
Fig. \ref{fig_max_geral_gamma}. On the other hand, the cusp-like behavior of
$\langle\overline{\mathcal{F}}\rangle$ occurs only at $T=0$, being smoothed 
out as we increase $T$. For higher values of $T$, the QCP is detected in this
case by a local maximum of $\langle\overline{\mathcal{F}}\rangle$ that occurs
exactly at the correct location of the QCP. 
For high enough $T$, though, this maximum 
is flattened to the point of becoming useless in spotlighting the QCP. 

The robustness of $\overline{\mathcal{F}}$ to detect the anisotropy QPT can be
traced back to its functional form and to the fact that 
it is exactly at $\gamma_c=0.0$
that $xx=yy$, with $|xx|<|yy|$ right before $\gamma_c$ and $|xx|>|yy|$ right after it (see Fig. \ref{fig4_apB} in the Appendix \ref{apB}). This feature is not changed
as we increase $T$ and it is the reason why the cusps of $\overline{\mathcal{F}}$
at $\lambda_c$ are not smoothed out or displaced 
as we increase $T$. We should also remark that 
the two cusps that we see in Fig. \ref{fig_med_geral_gamma} are not associated
with QPTs. They are a consequence of the functional form of 
$\langle\overline{\mathcal{F}}\rangle$ and to the following features
(see Fig. \ref{fig4_apB} in Appendix \ref{apB}). 
At the first cusp of $\langle\overline{\mathcal{F}}\rangle$, which occurs 
for $\gamma < \gamma_c$, $|zz|=|xx|$. Before this cusp we have 
$|zz|>|xx|$ and after it $|zz|<|xx|$. At the second cusp, which occurs 
for $\gamma > \gamma_c$, $|zz|=|yy|$. Before this second cusp we have 
$|zz|<|yy|$ and after it $|zz|>|yy|$. It is this exchange of the roles of
which two-point correlation function gives the greatest magnitude that causes 
those two peaks. Note that those peaks do not show up in 
$\overline{\mathcal{F}}$ because of its specific functional form, 
which implies that around those two locations it is only a function of either 
$yy$ or $xx$. Contrary to $\langle\overline{\mathcal{F}}\rangle$, 
there is no role for $zz$ in the functional form of 
$\overline{\mathcal{F}}$ about the locations of the two peaks seen for 
$\langle\overline{\mathcal{F}}\rangle$. Thus, the above discussion that explains
those peaks for $\langle\overline{\mathcal{F}}\rangle$ does not apply to
$\overline{\mathcal{F}}$ and hence there is no reason for those peaks to appear in
the functional behavior of $\overline{\mathcal{F}}$.

\section{Discussion} 

The present proposal to detect quantum critical points (QCPs) with finite temperature data should be analyzed under two aspects. First, looking at its theoretical side, the most important piece of information we must have access to 
in order to calculate the fidelities $\overline{\mathcal{F}}$ and 
$\langle\overline{\mathcal{F}}\rangle$ is the density matrix
describing a pair of qubits from the spin chain
[Eqs.~(\ref{rhoAB}) or (\ref{rhoAB2}), for instance]. The two-qubit 
density matrix is obtained after tracing out all but two qubits from the canonical ensemble density matrix describing the whole chain in equilibrium with a thermal bath at temperature $T$. This two-qubit density matrix is a function of  
one- and two-point correlation
functions and as such we must rely on analytical or numerical techniques to obtain
those correlation functions to have access to the two-qubit density matrix.

The traditional way of characterizing quantum phase transitions (QPTs), 
in particular at $T=0$, is based on the knowledge of those correlation functions
too. By studying their behavior as we change the 
system's Hamiltonian, or the behavior of quantities that are functions 
of them such as the magnetization or the magnetic susceptibility, 
we can detect QCPs by discontinuities 
in the $n$-th order derivative of those quantities that occur exactly at the QCPs. 
We can also employ quantum information theory based tools 
to detect QCPs, such as entanglement or quantum discord 
\cite{wu04,oli06,dil08,sar08,wer10,wer10b}. To apply these quantum information QCP 
detectors, we also need the correlation functions used in 
the traditional approach to characterize QPTs. The method we proposed in
Ref. \cite{pav23} and explored further here needs those correlation functions
too. 

However, some of these tools, such as the magnetization or 
magnetic susceptibility, 
may not
properly identify the correct spot of 
the QCP with finite $T$ data \cite{wer10b}. Other tools, such as the 
entanglement of formation \cite{woo98}, become zero at and around the QCP as we 
increase $T$, showing that they are useless in helping us in the identification
of the QCP after a certain temperature threshold \cite{wer10b}. The tremendous
success of quantum discord to spotlight QCPs at finite $T$ came to the forefront in 
Ref. \cite{wer10b}, where it was shown that 
for the XXZ model with no external field both QCPs are
detected by discontinuities in the derivatives of quantum discord that occur 
at the exact location of the QCPs, even as we increase $T$. The present 
teleportation based tools to detect QCPs have the same remarkable attributes of
quantum discord when detecting the QCPs for the XXZ model with no field
\cite{pav23}. However, a new theoretical feature sets them apart from
any known finite temperature QCP detector that is as robust as the quantum discord:
scalability as we increase the system's Hilbert space dimension.

Indeed, the evaluation of quantum discord is an NP-complete problem \cite{hua14}. 
Thus, the calculation of quantum discord is an intractable problem for 
high spin systems \cite{mal16}. On the other hand, the computational
resources that are needed to calculate the maximum mean and maximum 
average fidelities 
are not so demanding. The maximum average fidelity 
$\langle\overline{\mathcal{F}}\rangle$ is computed by
repeating for each one of the four sets of unitary
operations $S_k$ the calculation of the average fidelity as given by Eq.~(\ref{Flangle}). The computation of the latter is straightforward 
and can be scaled in an efficient way to an $N$-dimensional input state 
$|\psi \rangle$ \cite{gor06,pav23}.
The maximum mean fidelity $\overline{\mathcal{F}}$
is computed by repeating four times the maximization of Eq.~(\ref{Fbar}) over all input states $|\psi \rangle$(for each one of the four sets $S_k$ of unitary operations 
available to Bob). This optimization problem is much less demanding than
solving the optimization problem to determine the quantum discord or the 
entanglement of formation \cite{pav23}. All things being equal, the tools 
created in Ref. \cite{pav23} and further developed in this work to detect QCPs with finite $T$ data should rank among the most efficient, scalable, and robust tools
that are available in a theoretician tool box.

The second aspect under which the present proposal should be analyzed is its experimental interpretation and feasibility. Contrary to quantum discord, 
the teleportation based tools to detect QCPs here developed have a clear 
operational interpretation. The experimental steps needed to teleport a qubit,
namely, Bell state measurements and local unitary operations on single qubits, 
are clear and are not far from being implemented in spin-chain-like systems using 
state of the art techniques \cite{ron15,bra19,xie19,noi22,xue22,mad22,xie22}.
To experimentally 
determine Eqs.~(\ref{Fbar}) and (\ref{Flangle}), all we need to know 
is Bob's states at the end of several runs of the teleportation protocol 
using a representative sample of input states lying on the Bloch 
sphere as the states to be teleported from Alice to Bob.
Moreover, in order to experimentally obtain Bob's state once the 
teleportation protocol is implemented, we need to be able to measure the 
single spin density matrix describing Bob's qubit. In other words, we only 
have to experimentally obtain one-point correlation functions. 
There is no need to measure two-point correlation functions anymore.
Putting it differently, 
we can see the present proposal as a way to locally 
determine a QCP even when $T>0$. There is no need to globally
study the system in order to characterize its QPT with finite $T$ data. Another approach where only local measurements are enough to study QPTs 
at $T=0$ can be built using the quantum energy teleportation protocol
\cite{ike23a,ike23b,ike23c,ike23d} and its possible extension in 
the theoretical framework of quantum networks \cite{wan22,neu22,ike23e}.

It is also worth mentioning that from the 
experimental point of view, the time needed to implement all the steps of 
the teleportation protocol should be shorter than the time the system takes 
to return to equilibrium with the heat bath. The rate at which we
execute the teleportation protocol must be greater
than the relaxation rate of the system. We must also determine the state received
by Bob at the end of the teleportation protocol before it equilibrates once again with the heat bath. 

The theoretical computation of the relaxation time for 
an infinite spin chain is not trivial and lies beyond the scope of the present work. The relaxation time depends not only on the spin chain internal dynamics 
(its Hamiltonian) but also on how it interacts with the heat bath 
after a ``disturbance'' 
(the implementation of the teleportation protocol in our case). 
Experimentally, this relaxation time can be measured, for instance, 
by monitoring the magnetization of the system. In thermal equilibrium, 
the system’s magnetization has a definite value. When we perturb it, the magnetization changes. By monitoring the magnetization after the disturbance 
we can determine the time for the magnetization to get back to its equilibrium value. This time is the relaxation time, which is much easier to be measured than computed.

Furthermore, spin chains can be experimentally implemented on several different platforms. A few examples include quantum dots, quantum wells, 
and superconducting qubits. In GaAs quantum wells, 
for instance, we have at room temperature a relaxation time of a 
few nanoseconds \cite{ohn99}. In GaAs quantum dots 
the relaxation time of a spin-1/2 is measured to be around 50 $\mu s$ 
at $T\approx 20 mK$ \cite{han03} and in Ge/Si quantum dot arrays at 
$T \approx 5K$ we get a relaxation time around $10 \mu s$ \cite{zin10}. 
On the other hand, in silicon quantum dots at $T \approx 1K$ one can execute single and two-qubit gates in a time span shorter than $100 ns$
\cite{pet22}. This means that currently for silicon quantum dots and 
at low temperatures ($\approx 1 K$) we can, 
in principle, implement about one hundred gates before the system thermalizes. This is more than enough to implement  
the present proposal, which needs just a few gates at a given  
run of the teleportation protocol. We should also note that 
for superconducting qubits, we already have tens of qubits prepared
simultaneously with coherence times of the order of $100 \mu s$.
The time needed to implement single and two-qubit gates in this setup 
range between $10 ns$ to $100 ns$. This means that 
per coherence time we can implement  
between $10^3$ to $10^4$ gates \cite{gam17}.

\section{Conclusion}

We applied to several other models the teleportation based tools to detect
quantum critical points (QCPs) that were first presented in 
Ref. \cite{pav23}. 
We studied several spin-1/2 chains in the thermodynamic limit
(infinite number of spins). First we studied the XXZ model 
in an external longitudinal magnetic field and then the Ising model, the isotropic XX model, and the anisotropic XY model, all of them in 
external transverse magnetic fields. For all these models we investigated the performance of those tools to correctly detect the QCP at zero and finite temperature.\footnote{In the present work as well as in 
Ref. \cite{pav23} we have dealt with local models only. However,
the present quantum teleportation based tools to detect QCPs should
work equally well for non-local ones. This is true since what matters
most to the usefulness of the present tools is the fact that a QPT
induces a drastic change in the system's ground state. As such, the 
efficiency of the teleportation protocol should be affected as 
we cross the QCP irrespective of whether or not the interaction is local.} 

The key idea leading to the teleportation based tools to detect QCPs is the use
of a pair of spins from the spin chain as the entangled resource to 
implement the teleportation protocol. An external spin from 
the chain (the input state) is then teleported to one spin of that pair. 
We showed that the efficiency of the teleportation protocol depends crucially on 
which quantum phase we prepare the spin chain. At the QCP, we observed an abrupt
change in the efficiency of the teleportation protocol. The efficiency is 
quantified via the fidelity between the input state (Alice's qubit) and the 
output state at the end of the protocol (Bob's qubit).

For $T=0$ we verified that the maximum mean fidelity $\overline{\mathcal{F}}$ 
and the maximum average fidelity $\langle\overline{\mathcal{F}}\rangle$ have
a cusp or an inflection point exactly at the QCPs. For $T>0$ many of these cusps 
are smoothed out and both these cusps and the inflection points move away from the correct location of the QCPs. For finite $T$ these cusps and inflection points 
can be determined by studying the magnitudes of the first and second order derivatives of $\overline{\mathcal{F}}$ and $\langle\overline{\mathcal{F}}\rangle$.
The magnitudes of these derivatives become very large around the location of the QCPs. Below a certain temperature threshold, the locations of the extrema for those
derivatives lie in a straight line and by extrapolating to zero temperature we
can predict the correct values of the QCPs.

The results of Ref. \cite{pav23} and the ones shown here imply 
that the teleportation 
based tools to detect QCPs have the same 
important characteristics of quantum discord \cite{wer10b}, 
one of the most reliable QCP detector for finite $T$. Both quantum discord and
the teleportation based tools studied here can be applied without the knowledge of the order parameter related to the QPT and are very robust to 
temperature increases. In addition to that, the present tools 
have two important characteristics not shared with quantum discord \cite{pav23}. 
First, they have a direct experimental meaning while quantum discord does not. 
Also, the computational resources that we need 
to theoretically calculate them is much less demanding than what is required to compute quantum discord. This fact allows us to scale the present 
tools to high spin systems.

Finally, looking at Figs. \ref{fig_max_geral6}, \ref{fig_max_geral12}, \ref{fig_max_geral_xy}, \ref{fig_max_geral_xy2}, 
\ref{fig_max_geral_xy3}, and \ref{fig_max_geral_gamma}, we observe that for 
each model and for each phase transition the behavior of 
$\overline{\mathcal{F}}$ is unique (a similar analysis applies to
$\langle\overline{\mathcal{F}}\rangle$). In other words, the fingerprint
of a phase transition and its underlying model is unique. The functional 
behavior of $\overline{\mathcal{F}}$ as we change the tuning parameter of the Hamiltonian and drive the system across the QCP is specific for each model.
In this sense, by studying $\overline{\mathcal{F}}$ we can not only detect
a QCP but also pinpoint the underlying model that led to that phase transition.

\begin{acknowledgments}
GR thanks the Brazilian agency CNPq
(National Council for Scientific and Technological Development) for funding and 
CNPq/FAPERJ (State of Rio de Janeiro Research Foundation) for financial support
through the National Institute of Science and Technology for Quantum Information.
GAPR is grateful to the São Paulo Research Foundation (FAPESP)
for financial support through the grant 2023/03947-0.
\end{acknowledgments}

\appendix

\section{Correlation functions for the XXZ model in an external field}  

\label{apA}

\begin{figure}[!ht]
\includegraphics[width=8cm]{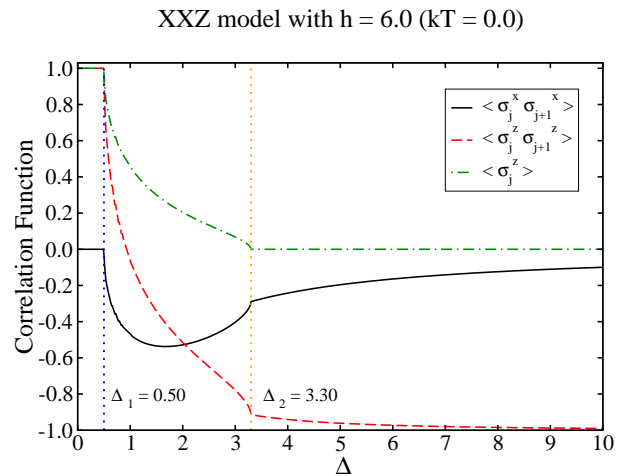}
\caption{\label{fig1_apA}(color online) One- and two-point correlation 
functions as a function of the tuning
parameter $\Delta$ with external magnetic field $h=6.0$. 
All data were computed in the thermodynamic limit and at $T = 0$. 
The dotted lines mark the two QCPs for this model.}
\end{figure}

\begin{figure}[!ht]
\includegraphics[width=8cm]{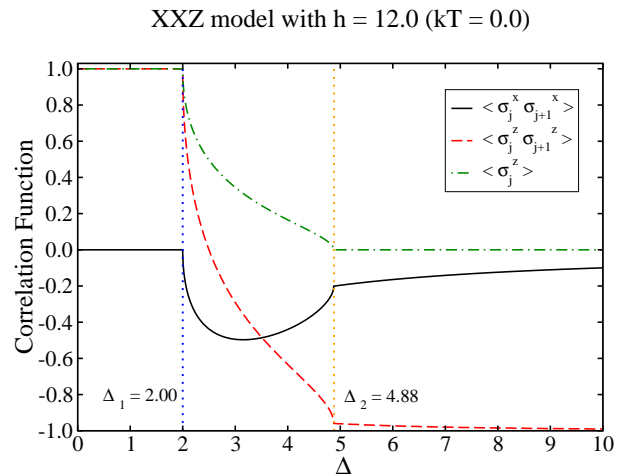}
\caption{\label{fig2_apA}(color online) Same as Fig. \ref{fig1_apA} but with
$h=12.0$.}
\end{figure}

The Hamiltonian describing the $XXZ$ model 
in
a longitudinal external magnetic field is given by Eq.~(\ref{Hxxz}). The solution to this model for 
arbitrary $T$ is given by Refs. \cite{klu92,bor05,boo08,tri10} and this solution
was adapted to the present purposes in Ref. \cite{wer10b}. At the absolute zero temperature, the functional behavior of the non-null correlation functions 
is given by Figs. \ref{fig1_apA} and \ref{fig2_apA}.

\begin{figure}[!ht]
\includegraphics[width=8cm]{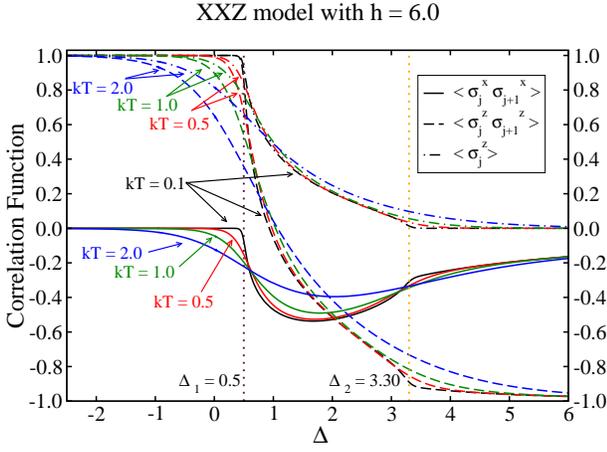}
\caption{\label{fig3_apA}(color online) One- and two-point correlation 
functions as a function of the tuning
parameter $\Delta$ with external magnetic field $h=6.0$. 
All data were computed in the thermodynamic limit for several values of $T > 0$. 
The dotted lines mark the two 
QCPs for this model.}
\end{figure}

\begin{figure}[!ht]
\includegraphics[width=8cm]{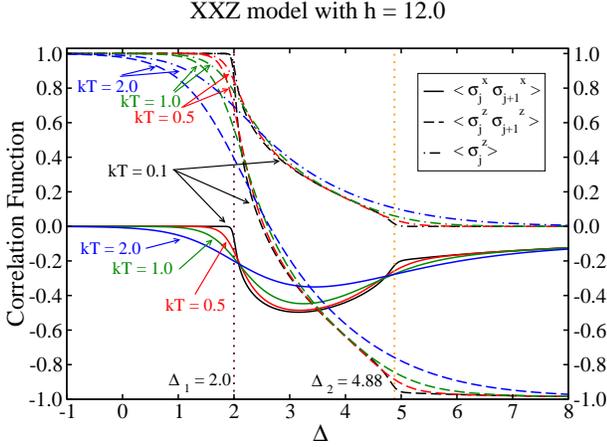}
\caption{\label{fig4_apA}(color online) Same as Fig. \ref{fig3_apA} but with
$h=12.0$.}
\end{figure}

For finite $T$, the correlation functions as a function of $\Delta$ are given by Figs. \ref{fig3_apA} and \ref{fig4_apA}.

\section{Correlation functions for the XY model subjected to an external field}
\label{apB}

The Hamiltonian describing the $XY$ model subjected to a transverse external magnetic field is given by Eq.~(\ref{Hxy}). This model was solved for 
arbitrary $T$ in Refs. \cite{lie61,bar70,bar71}. Using the present notation,
a step-by-step description of this solution can be found in Ref. \cite{wer10b}.
Note that two typos should be taken into account when consulting Ref. \cite{wer10b}.
The Hamiltonian for the XY model there presented lacks an overall $1/2$ factor and 
the expression for $\langle \sigma_j^z\rangle$ should be multiplied by $-1$.

\begin{figure}[!ht]
\includegraphics[width=8cm]{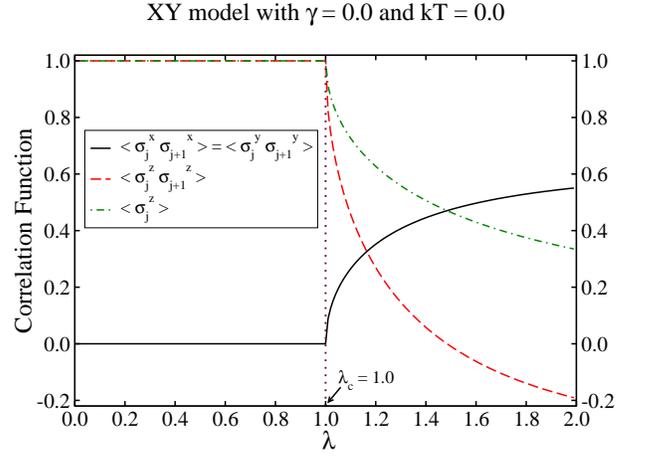}
\caption{\label{fig1_apB}(color online) One- and two-point correlation 
functions as a function of $\lambda$, the inverse strength of the field.
Here $\gamma=0.0$. All data were computed in the thermodynamic limit and at 
$T = 0$. The dotted vertical line marks the QCP for this model.}
\end{figure}

\begin{figure}[!ht]
\includegraphics[width=8cm]{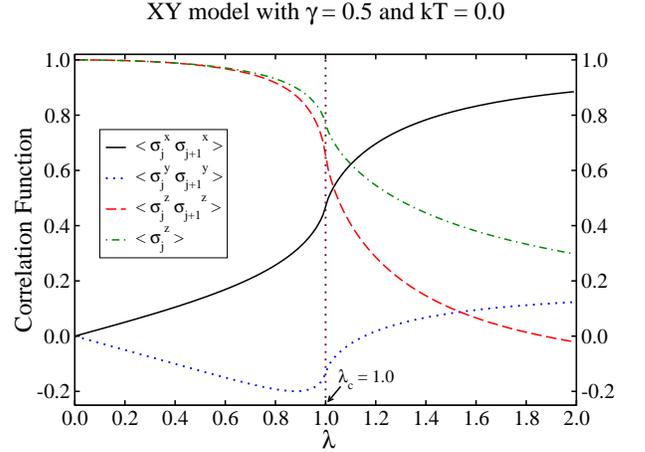}
\caption{\label{fig2_apB}(color online) Same as Fig. \ref{fig1_apB} but with
$\gamma=0.5$.}
\end{figure}

\begin{figure}[!ht]
\includegraphics[width=8cm]{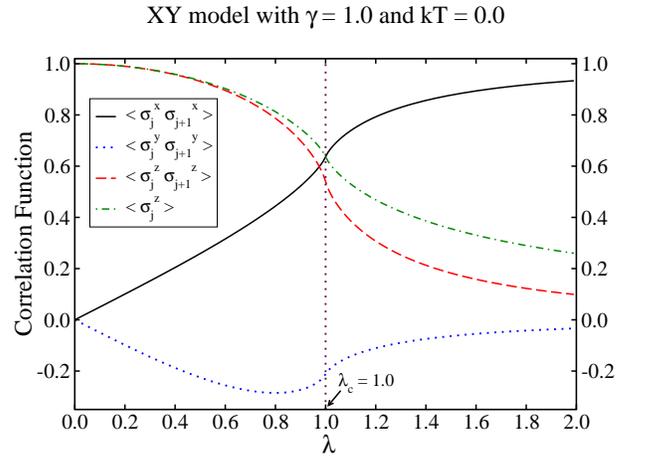}
\caption{\label{fig3_apB}(color online) Same as Fig. \ref{fig1_apB} but with
$\gamma=1.0$.}
\end{figure}

In Figs. \ref{fig1_apB}, \ref{fig2_apB}, and \ref{fig3_apB} 
we plot for $T=0$ the non-zero correlation functions for this model as a function of
$\lambda$ for the three values of $\gamma$ employed in the main text. Note 
that the case where $\gamma=1.0$ is the transverse Ising model.  

\begin{figure}[!ht]
\includegraphics[width=8cm]{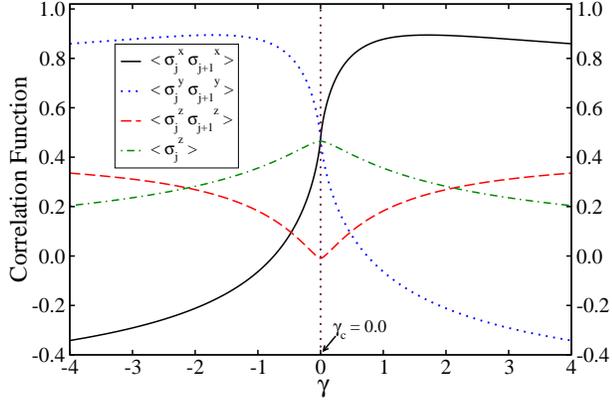}
\caption{\label{fig4_apB}(color online) One- and two-point correlation 
functions as a function of $\gamma$, the anisotropy parameter.
Here $\lambda=1.5$. All data were computed in the thermodynamic limit and at 
$T = 0$. The dotted vertical line marks the QCP for this model.}
\end{figure}

For $T=0$ and fixing $\lambda=1.5$, we show in Fig. \ref{fig4_apB} 
the relevant correlation functions as we change the anisotropy parameter 
$\gamma$.

The respective curves for $T>0$ have the general trends of the 
$T=0$ curves and we will not show them here. Similarly to what we see for the
finite $T$ curves of the XXZ model (see the Appendix \ref{apA}), 
as we increase the temperature the kinks are smoothed out and 
displaced from their locations at $T=0$. Also, 
the magnitudes of the first and second order derivatives of the 
correlation functions at the QCPs decrease and are displaced from their 
$T=0$ locations.

\end{document}